\documentclass{article}

\usepackage{arxiv}

\usepackage[utf8]{inputenc} % allow utf-8 input
\usepackage[T1]{fontenc}    % use 8-bit T1 fonts
\usepackage{hyperref}       % hyperlinks
\usepackage{url}            % simple URL typesetting
\usepackage{booktabs}       % professional-quality tables
\usepackage{amsfonts}       % blackboard math symbols
\usepackage{nicefrac}       % compact symbols for 1/2, etc.
\usepackage{microtype}      % microtypography
\usepackage{lipsum}		% Can be removed after putting your text content
\usepackage{graphicx}
\usepackage{natbib}
\usepackage{doi}
\usepackage{subfiles}
\usepackage{float}

\begin{document}

\date{}

\title{Sirius: Visualization of Mixed Features as a Mutual Information Network Graph}

\author{\href{https://orcid.org/0000-0002-7826-3500}{\includegraphics[scale=0.06]{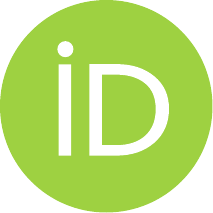}\hspace{1mm}Jane Lydia Adams\thanks{Data Visualization Lab, Khoury College of Computer Sciences, Northeastern University, Boston, Massachusetts.}}
\\\texttt{adams.jan@northeastern.edu}
\And
\href{https://orcid.org/0000-0001-8809-7991}{\includegraphics[scale=0.06]{orcid.pdf}\hspace{1mm}Todd F. DeLuca\thanks{University of Vermont Complex Systems Center, Burlington, Vermont.}}
\\\texttt{Todd.DeLuca@uvm.edu}
\And
\href{https://orcid.org/0000-0002-9857-2845}{\includegraphics[scale=0.06]{orcid.pdf}\hspace{1mm}Christopher M. Danforth\textsuperscript{†}}
\\\texttt{Chris.Danforth@uvm.edu}
\And
\href{https://orcid.org/0000-0003-1973-8614}{\includegraphics[scale=0.06]{orcid.pdf}\hspace{1mm}Peter Sheridan Dodds\textsuperscript{†}}
\\\texttt{Peter.Dodds@uvm.edu}
\And
Yuhang Zheng\thanks{Massachusetts Mutual Life Insurance Company, Springfield, Massachusetts.}
\\\texttt{YZheng83@MassMutual.com}
\And
Konstantinos Anastasakis\textsuperscript{‡}
\\\texttt{KAnastasakis26@MassMutual.com}
\And
Boyoon Choi\textsuperscript{‡}
\\\texttt{BChoi19@MassMutual.com}
\And
Allison Min\textsuperscript{†}
\\\texttt{AMin13@massmutual.com}
\And
Michael M. Bessey\textsuperscript{‡}
\\\texttt{MBessey@MassMutual.com}
}

\maketitle


\begin{abstract}
\begin{figure}[H]
\centering
\includegraphics[width=\textwidth]{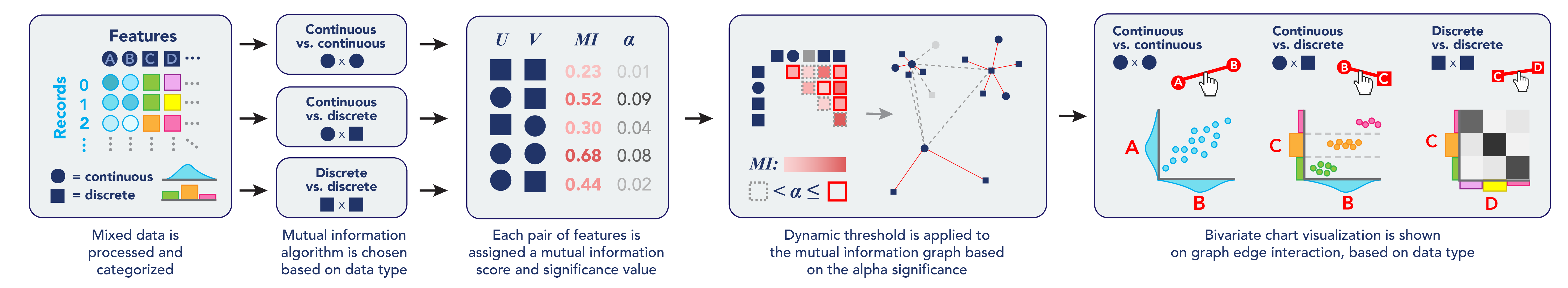}
\caption{\label{teaser}
           Flowchart describing the data processing and visualization of the mutual information feature network using Sirius. For the purpose of this explanatory graphic: circles represent continuous features, and squares represent discrete features.
}
\end{figure}
  Data scientists across disciplines are increasingly in need of exploratory analysis tools for data sets with a high volume of features of mixed data type (quantitative continuous and discrete categorical). We introduce Sirius, a novel visualization package for researchers to explore feature relationships among mixed data types using mutual information. The visualization of feature relationships aids data scientists in finding meaningful dependence among features prior to the development of predictive modeling pipelines, which can inform downstream analysis such as feature selection, feature extraction, and early detection of potential proxy variables. Using an information theoretic approach, Sirius supports network visualization of heterogeneous data sets (consisting of continuous and discrete data types), and provides a user interface for exploring feature pairs with locally significant mutual information scores. Mutual information algorithm and bivariate chart types are assigned on a data type pairing basis (continuous-continuous, discrete-discrete, and discrete-continuous). We show how this tool can be used for tasks such as hypothesis confirmation, identification of predictive features, suggestions for feature extraction, or early warning of data abnormalities. The accompanying website for this paper can be accessed at \textit{https://sirius.universalities.com/}. All code and supplemental materials can be accessed at \textit{https://osf.io/pdm9r/}.
\end{abstract}


\begin{figure}[H]
  \includegraphics[width=\textwidth]{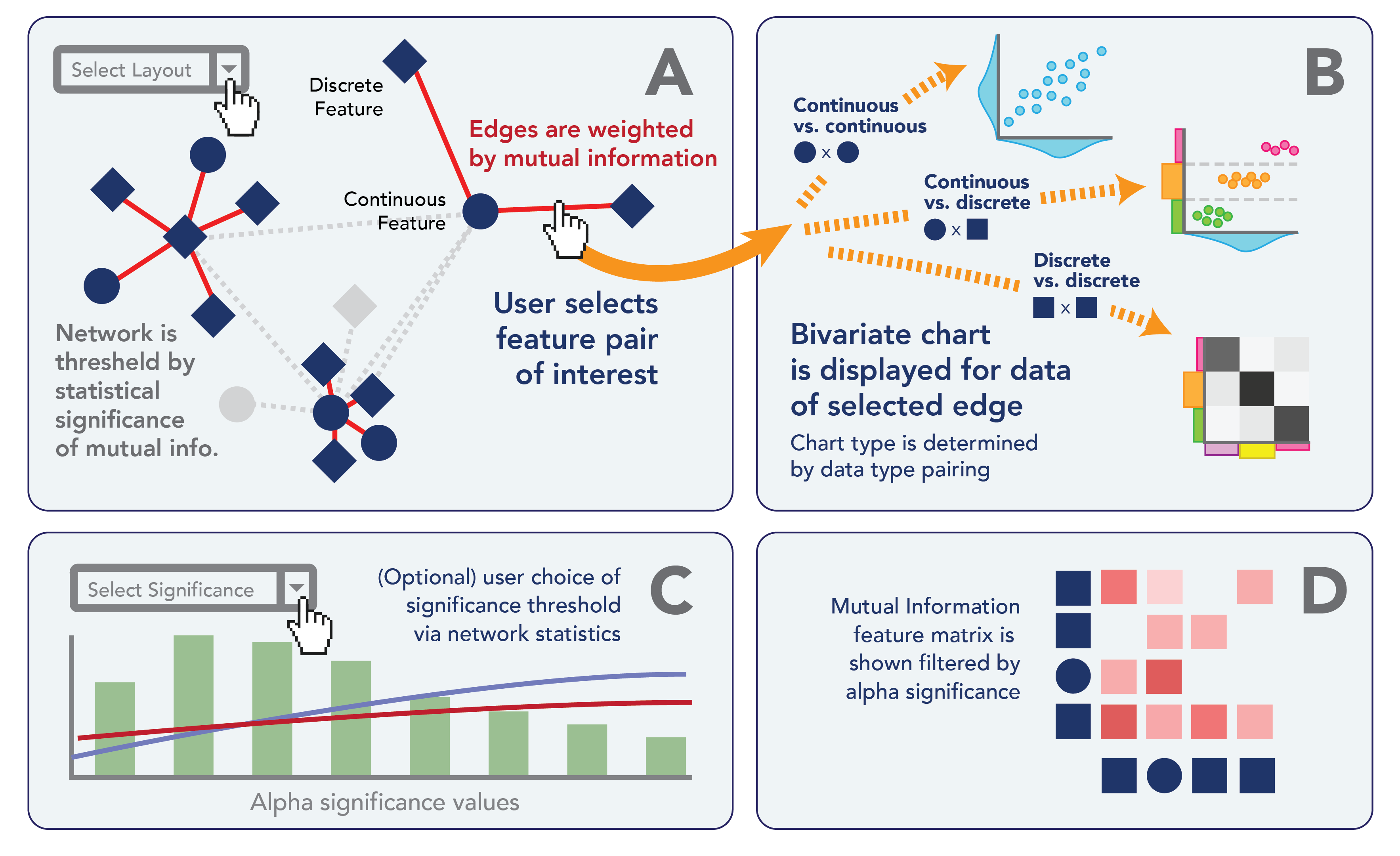}
  \caption{We show the four main components of the Sirius exploratory analysis dashboard: \textbf{A) The mutual information feature network}, in which discrete and continuous features are both represented as nodes in a graph with edges weighted by mutual information and sparsified according to their statistical significance. \textbf{B) The bivariate chart view}, which renders a chart showing the raw data for 2 features of interest based on the selected edge from the network graph in (A). A major contribution of this system is the automatic selection of one of three visual encodings, informed by the data types chosen. \textbf{C) The alpha significance explorer}, which allows users to review information related to the statistical significance of mutual information edges in the network graph and, optionally, to customize the significance threshold applied to the network. \textbf{D) The matrix representation} of the feature network shown in (A), shown as a heatmap where color corresponds to the mutual information values and features are sorted by information gain using hierarchical clustering.}
  \label{fig:expl}
\end{figure}

\section{Introduction}
Statistical analysis provides robust mathematical insight for a wide range of domain applications, including finance \cite{Griebel:2010:DWI}, marketing \cite{Russell:2000:ACC,Raeder:2011:MBA}, genomics \cite{Langfelder:2008:WGC,Nam:2007:CSV,Pyne:2009:AHD}, chemistry \cite{Dordrecht:1984:MDA,Yapeng:2017:SCP,McCarthy:2004:AML}, environmental science \cite{Warton:2010:RSE}, and social interaction \cite{Gao:2015:PCH,Gotz:2014:DVA,Shuman:2014:EFS}. Advances in computational power and storage, coupled with a rapid uptick in data collection and exchange, have caused massive increases in the volume of data collected: by some estimates, 90\% of all current data was created in the past few years \cite{Menon:2015:STA, Al-Jarrah:2015:BDA}. This has engendered the development of powerful machine learning tools which support imputation, clustering, and prediction \cite{Fahad:2014:SCA,Ward:2013:UDS,Chen:2014:BDS}. When designing exploratory systems for high dimensional data, researchers are often left with the choice of aggregating the data using dimension reduction, having many views to show raw data, or presenting summary statistics. Each of these methods presents a host of issues for the researchers exploring the data. Our proposed visualization with Sirius helps to minimize these problems by providing researchers a direct and accessible way to accurately explore a smaller subset of raw data, thereby increasing researchers' ability to find insights in large multidimensional datasets. The purpose of Sirius as a visualization tool is to inform subsequent hypothesis generation, clustering, feature selection, imputation and predictive modeling in a data science pipeline. 

Additionally, there are growing ethical concerns among researchers and the general public about `black box' models, which can obscure inadvertent statistical biases in the machine learning pipeline and lead to disparate real-world impacts (for example: lower home appraisal values or higher risk assessments) \cite{Zafar:2017:FBD,Feldman:2015:CRD}. Simply excluding `protected class' data, such as race or gender, from model inputs is not sufficient to offset disparate impacts when other features act as proxy variables \cite{Wang:2018:DDI,Skeem:2016:RRR}. For example: population incidence of Sickle cell anemia varies significantly by race, so a model which places highest feature importance on the presence of that disease may inadvertently be encoding systemic racism into mortality risk assessments, which could in turn result in higher insurance premiums for certain subpopulations. This problem is in part due to the obfuscation of raw data (e.g., through summary statistics, feature projection, or complex neural network architectures) from data scientists during the exploration and development phases of the statistical modeling pipeline. Exploratory Data Analysis (EDA) preserves data scientists' access to record level data, while addressing the growing need to holistically explore a large number of complexly related features, thereby preserving the nature of individual data points within the visual encoding \cite{Behrens:1997:PPE}. This is where we have derived the name Sirius for our exploratory analysis package: Sirius, the brightest star in the night sky, provided navigational support for sailors, just as we hope to provide directional insight for wayward data scientists \cite{brittanica_sirius}.

To address these challenges of exploring feature dependence of mixed data types, we introduce this package for generating exploratory visualization dashboards for high-dimensional mixed data through the integration of existing state of the art algorithms and tools for computing mutual information, generating network graphs, sparsification of matrices, and visualization into a single tool. A primary motivating criteria for the development of Sirius is to support mixed data types (categorical, ordinal, binary, and continuous). We contrast this with other systems which do not support mixed data at all and only allow feature dependence comparisons across entirely continuous or entirely discrete data sets, or do not support mixed data without transformations such as binning. 

Mutual information is a measure of the decrease in entropy when making a prediction about one variable with the addition of data about another variable. For example: Predicting a person's pregnancy status in a healthcare setting based on their height may yield results hardly better than random guessing; but the gender of a patient (a discrete categorical variable) or their Human Chorionic Gonadotropin (hCG) blood test results (a continuous variable) may confer additional predictive power. Therefore, we might expect 'pregnancy', 'sex', and 'hCG' features to all share high mutual information with one another. Further details about mutual information can be found in Section \ref{calc_mi}. The result is an interactive network graph of feature dependence weighted by mutual information, with charts of 2-dimensional feature relationships shown upon interaction (Fig.~\ref{teaser}). These relationships are further filtered in the network graph using a graph-aware sparsification approach called backbone thresholding, which leverages the statistical significance of edges in relation to feature nodes in order to dynamically remove insignificant edges from the visual representation. The mutual information feature network aids exploratory analysis of feature-level relationships by narrowing the scope of visual comparison among a subset of variables through statistical methods. Simultaneously, our system preserves access to record-level data through the rapid rendering of bivariate charts showing individual data points across the paired features of interest. Sirius was developed through an interactive design process in close collaboration with data analysts using mixed-type data, and validated through user feedback and subsequent hypothesis generation, feature selection, and data re-collection efforts on several data sets (4 sample data sets included in the public repository, along with 2 additional data sets in real-world contexts related to human health and co-morbidity modeling).

\textbf{Contributions:} In this paper, we contribute a novel tool for handling mixed data types in a high-dimensional exploratory data analysis pipeline without binning or projection. Sirius is a proof of concept implementation for this novel visual rendering approach, which relies on automated selection of 1) mutual information algorithm and 2) subsequent bivariate chart type, based on data type. Sirius is one of the first systems to handle both discrete and continuous data for extremely high-dimensional data, and the first to not discretize continuous data.  By keeping continuous data continuous and not projecting data points into lower-dimensional space, we enable data analysts to retain an un-obfuscated view of raw record-level data, while providing an overview of feature dependencies in a graph-aware manner.

\section{System Design Criteria}\label{objectives}
In our user interviews, we identified several high-level objectives, which we then distilled into the following five specific criteria for our system, which contrast it with state-of-the-art existing research in this problem space:

\begin{itemize}
    \item \textbf{C1: Support mixed data types} Due to the varied nature of data in contemporary analysis pipelines, our users require a system for understanding feature relationships that allows comparisons between discrete and continuous data types. Therefore, existing methods that compare only continuous features or only discrete features are insufficient. The system must support comparisons among 1) continuous-continuous, 2) continuous-discrete, and 3) discrete-discrete pairings of features. This is informed by our user interview objective to support EDA of high numbers of features of mixed continuous and discrete data types, as these large-scale heterogenous data sets are common in a wide range of real-world data analysis tasks.
    \item \textbf{C2: Show dimension space and data item space} Since users are interested both in viewing abstract feature dependence and underlying raw data, we aspire to Schniederman’s Mantra: Overview first, zoom and filter, details on demand \cite{mantra}. Therefore we framed our system as predominantly a view bifurcated into A) \textit{dimension space} (showing feature relationships) and B) \textit{data item space} showing specific data points for a subset of dimensions. This allows for rapid task switching between macro-scale, 'big picture' feature relationships; and micro-scale, record-level exploratory analysis, wherein specific subpopulation and individual data can be accessed visually.
    \item \textbf{C3: Do not distort raw data} Methods abound for feature selection (subsetting, e.g., based on importance scores from a random forest), extraction (e.g., combining into basis vectors), and projection (e.g., manifold learning); however, when talking with users, these methods were not of necessity for this step of exploratory analysis. Rather, users wanted to be able to view raw underlying data \textit{before} potentially mathematically obfuscating important information such as outliers, dirty data, or unexpected dependence. Therefore, we set a constraint that all data item space visualizations must offer as granular a view as possible into the true nature of the data distributions. This also allows users to highlight potential 'proxy variables' or features with implications for privacy and equity. Unexpected feature relationships can sometimes portend disparate impacts for certain populations, or deanonymization vulnerabilities in sensitive data. For example, low appraisal values of homes in certain zip codes may warrant investigation of potential redlining effects. An innocuous ``ID number" feature intended to anonymize protected patient health data, when shown to be associated with a hospital admittance source, can drastically narrow a search space for nefarious actors seeking to deanonymize data. Bringing these unforeseen associations to the attention of decision-makers can provide vital insight for policy-making and have significant consequences for stakeholders.
    \item \textbf{C4: Ensure system modularity} Given that this system addresses one portion of an exploratory visual analytics pipeline, we want components of the system to be \textit{reusable, distinct,} and \textit{extensible}. What this means in practice is that data processing functions can be \textit{individually utilized} in other contexts for rapid prototyping of alternate views; encodings are \textit{separable}, e.g., visual components can be extracted and embedded elsewhere while retaining interactivity; and \textit{extensible} such that end users, many of whom are also programmers, might easily make and submit domain-specific or added functionality modifications to the code base for posterity.
    \item \textbf{C5: Open source project} Making code open source allows for greater transparency; can improve quality of research and saves valuable time; encourages two-way informational exchange between developers and end users; and aligns with educational and accessibility goals of open access science more broadly.
\end{itemize}

\section{Related Work}

\begin{table}[]
\begin{tabular}{rlllll}
\textbf{System} & \textbf{C1} & \textbf{C2} & \textbf{C3} & \textbf{C4} & \textbf{C5} \\ \hline
VEIMR Viewer \cite{Bernard2014} & \checkmark & \checkmark & - & - & - \\
HCE \cite{seo_rank-by-feature_2005} & - & \checkmark & \checkmark & - & \checkmark \\
INFUSE \cite{infuse} & - & - & - & - & \checkmark \\
MDS Viewer \cite{dimension_projection-yuan} & - & \checkmark & - & - & - \\
RFG Viewer \cite{Turkay2012} & - & - & - & - & - \\
ParSets \cite{Parallelsets} & \checkmark & \checkmark & - & - & \checkmark \\
CComViz \cite{Zhou2009VisuallyCM} & - & - & - & - & - \\
ScagExplorer \cite{scagexplorer} & - & \checkmark & \checkmark & - & - \\
XCluSim \cite{xclusim} & - & \checkmark & - & - & - \\
VICTOR \cite{victor} & - & \checkmark & \checkmark & \checkmark & \checkmark \\
Sirius & \checkmark & \checkmark & \checkmark & \checkmark & \checkmark \\
\multicolumn{1}{l}{} &  &  &  &  & 
\end{tabular}
\caption{We compare Sirius to other tools for exploratory analysis of high-dimensional data along our criteria \textbf{C1} Support mixed data types, \textbf{C2} Show dimension space and data item space, \textbf{C3} Do not distort raw data, \textbf{C4} Ensure system modularity, and \textbf{C5} Open source project. We show that while existing methods do address some of these criteria, our users were in need of a system that met all of the requirements, and therefore Sirius is an important addition to the ecosystem of tools for high-dimensional exploratory analysis. An interactive version of this chart is available on the project website, \textit{https://sirius.universalities.com}.}
\end{table}

The data visualization subfield of Exploratory Data Analysis (EDA) for high-dimensional data has been established as a result of the encoding challenges of increasing the number of dimensions in a data set. Traditional visualization methods are generally quite resilient to \textit{vertically scalability} (increasing the number of record); the challenge for visual representation then is primarily in computational power (for example, a browser's memory might struggle to support rendering of thousands of circle glyphs in a scatter plot on a web page). However, \textit{horizontal scalability} (increasing the number of features) when visualizing data for exploratory analysis presents a challenge for traditional visual encoding methods: while a data set with 4 features $f$ might visually encode $f_1$: x-position, $f_2$: y-position, $f_3$: point size and $f_4$: color for each record in a scatterplot, increasing the number of features to 100 is not simply a question of increasing visual complexity to add z-position, shape, texture, time, transparency, sonification, etc. encodings up to $f_{100}$. Instead of continuing this line of ever-growing encoding complexity, it is vitally necessary to develop novel applications of visual or statistical summary methods for analysis of feature dependence in a visual context.

Including all features in a prediction task can be computationally expensive, and the addition of unrelated features can sometimes inhibit predictive accuracy due to overfitting and the 'curse of dimensionality' wherein high-dimensional spaces are increasingly sparse  \cite{Bertini:2011:QMH}. For example, random forest regression models can generate feature importance rankings; subsequent modeling with the exclusion of low-importance features can improve model accuracy, but this iterative approach can be time- and resource-prohibitive. Additionally, related features could be grouped together in a lower-dimensional visualization, as through Principal Component Analysis (PCA). For example, in a healthcare setting, there may be many features related to body systems, such as blood sugar, cardiovascular status, or liver health. Quickly identifying all such groupings can allow data scientists to extract a single new vector to represent all features in a given system and more easily generate meta-analyses of interacting body systems. We developed Sirius to support analysts in the identification of groupings of related features to inform subsequent analytic transformations. \textbf{Of critical importance for end users was the ability to include mixed data types (that is, continuous \textit{and} discrete) in an Exploratory Data Analysis visualization; a criteria that is not sufficiently met by existing tools.} 

One approach for EDA of a limited number of features is small multiples \cite{maceachren_exploring_2003}. By tiling features across rows and columns, all permutations of paired feature charts can be viewed simultaneously. While this system originated with the scatterplot matrix, in its current form the pair plot is advantageous in that it supports visualization of heterogenous data types, by automatically selecting appropriate charts for each feature pairing (e.g., scatter plots for continuous-continuous pairings, heatmaps or conditional bar charts for discrete-discrete pairings, and violin plots or ridgeline plots for discrete-continuous pairings). This supports our {C1: Support mixed data types}. However, pair plots quickly diminish in visualization efficacy as the number of features grows, because of the difficulty of fitting every permutation of feature pair plot into a single visual space \cite{scagnostics}. A notable example of this is the \textit{Hierarchical Clustering Explorer (HCE)}, which uses hierarchical agglomerative clustering (HAC) to group features by similarity \cite{seo_rank-by-feature_2005}. The HCE dashboard then generates similarity matrices of features. Matrices are often visually represented as heatmaps which are colored according to the summary statistics of feature comparisons. These can accommodate higher numbers of feature comparisons in the visual space due to their summational behavior. Sirius leverages mutual information as a distance measure through automatic selection of the correct algorithm for the data types of interest. This is not the case with Hierarchical Clustering Explorer ~\cite{seo_rank-by-feature_2005}, which does offer mutual information as a possible distance measure, but only computes distance measures across continuous features.

Mutual information is a way of evaluating dependence of two features by measuring entropy reduction. If knowing the response to one feature for a given record increases one's accuracy of predicting that record's value for another feature, it is likely that these two features share high mutual information. In the mutual information feature network technique we present here, features with high mutual information would be connected in the resulting network graph. Conversely, if knowing the response to one feature does not significantly improve prediction of a second feature, those features are likely relatively independent and will have a low mutual information. In the Sirius application, an edge between these two unrelated features' nodes will likely not be displayed in the graph. Compare this to the research area of Scagnostics, in which SPLOMs (scatterplot matrices) are generated based on bivariate plot characteristics such as: Outlying, Skewed, Clumpy, Convex, Skinny, Striated, Stringy, Straight, and Monotonic \cite{scagexplorer, scagnostics, tukey1977exploratory}. Mutual information is another such tool for measuring a relationship between two features, but it is not limited to continuous data types, as scagnostics are.

\label{assoc_corr}
Two prominent feature visualization methods employing network graph visualizations for understanding dimension relationships in tabular data include \textit{Association Networks} and \textit{Correlation Networks}. 

\textbf{Association networks} are commonly employed in market basket analysis, in which edges are drawn between items if purchasing one item increases the likelihood of purchasing a different item, a weight referred to as the `lift', derived from the conditional probabilities of purchasing two items in the same transaction \cite{Agrawal:1993:MAR}. Thus, these association networks are directed weighted networks, and are classically defined for boolean data types only. The resultant network visualizations show feature groupings among products which are frequently purchased together, and might provide actionable insight for marketers looking to target potential consumers of a certain product. 

\textbf{Correlation networks}, popularized by \textit{Weighted Gene Correlation Network Analysis (WGCNA)}, are commonly applied in genomics research \cite{Langfelder:2008:WGC}. Large network graphs are generated in which nodes represent genes, and edges are drawn according to co-expression correlation across cell lines. Data types for this kind of visualization are classically all continuous values, such as gene expression probabilities \cite{Niemira:2019:MSS}.

These kinds of feature networks, in high-dimensional settings, can often become extremely dense and difficult to navigate during exploratory analysis. Both association networks and correlation networks also do not support comparisons across heterogeneous data types. The key value propositions of the Sirius mutual information feature network method described in Section \ref{network_methods} are 1) backbone sparsification, which simplifies the exploratory space according to the statistical significance of similarities among features, and 2) comparison among and between continuous and discrete data types, as described in section \ref{calc_mi}.

Sirius enables quick visualization of information theoretic feature projection network graphs coupled with 2-feature plots of feature pairs of interest. This is consistent with other related work such as: \cite{dimension_projection-yuan}, which refers to these as "data item plots"; and Hierarchical Clustering Explorer \cite{seo_rank-by-feature_2005}. In both of these systems, all record-level plots are scatter plots, because the only feature data types supported by the systems are continuous. In Sirius, these charts could be beeswarm plots, heatmaps, or scatter plots, depending on the data type pairing for that specific comparison.

We choose a network graph visualization over a heatmap matrix as our primary encoding of feature similarities due to the large, sparse nature of the network. Sirius provides adjustable backbone sparsification of feature network graphs. The sparsification contribution is important to note, as matrix heatmaps can quickly grow quite large, while similarly, network graphs often fall prey to the famous "hairball" issue of extremely dense edge rendering. The method of context-aware sparsificiation is described in detail in Section \ref{network_methods}.

\section{System}
We couch our system within the technique classification, visualization pipeline, and user interaction paradigms of existing high-dimensional analysis tools. We then review the specific views of the dashboard visualization.

\begin{figure*}[htbp]
\centering
\includegraphics[width=15cm]{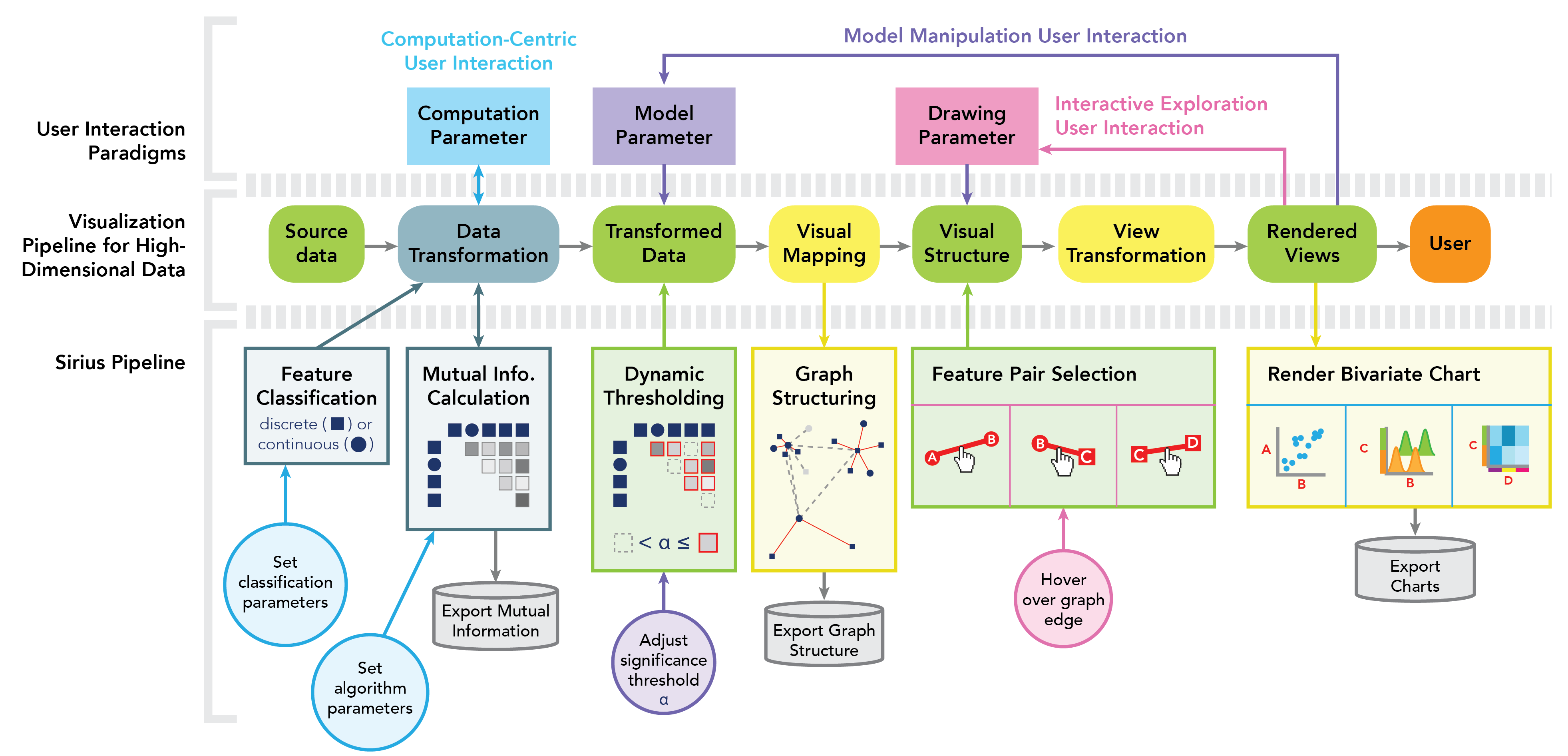}
\caption{\label{related_context}
           We situate Sirius within the contexts of 1) the visualization pipeline for high-dimensional data specified in \cite{Bertini:2011:QMH} and \cite{liu} and 2) the subsequent user interaction paradigms identified in \cite{liu}.
}
\vspace{-14pt}
\end{figure*}

\subsection{Technique Classification}
Keim et al. describe in \cite{keim-infovis} a classification for information visualization techniques with 3 characterizations: 1) \textit{Data type to be visualized}, 2) \textit{Visualization technique}, and 3) \textit{Interaction and distortion technique}. 

\textbf{C1: Support mixed data types} informs the characterization of Sirius as "Multi-dimensional", such as relational tables, as the \textit{data type to be visualized}, with emphasis on the additional specification that not only is the data high-dimensional, but also heterogeneous in nature.

Our motivational criteria \textbf{C2: Show dimension space and data item space} and \textbf{C3: Do not distort raw data} support our choice for \textit{Visualization technique} in this classification scheme as "Standard 2D/3D display", such as bar charts and x-y plots, much like \cite{Polaris}. This is an important distinction from tools like \cite{dimension_projection-yuan} and \cite{Turkay2012} which use MDS to generate (extract) representative factors based on feature similarity. While these methods of feature extraction are useful in predictive modeling pipelines, we emphasize that raw, non-projective data exploration is a critical first step in visual analytics for our users, as we have detailed in Section \ref{objectives}. Similarly, we distinguish our visualization needs to exceed the limits of other visualization techniques, namely: geometrically-transformed displays such as parallel coordinates; iconic display; dense pixel display; and stacked display, as many of these techniques are either limited in their ability to render high-dimensional data, or restricted to homogenous (e.g., only continuous) data types.

\subsection{Visualization Pipeline and User Interaction Paradigms}
In their survey of Advances in Visualizing High-dimensional Data, Liu et. al. describe the visualization pipeline for high-dimensional data and specify user interaction paradigms that take place at various steps of the process. We enumerate the pipeline of Sirius in this context, as diagrammed in Fig. \ref{related_context}, as such: \textbf{Source Data} Our tool as built is set up to take in tabular data from a csv file; however, due to the modular structure of the package, it is possible to interface with any tabular data source which can be accessed using Pandas. \textbf{Data Transformation} User modifications made to this step are classified by \cite{liu} as "Computation-Centric" user interactions: specifically, they are computation parameters that determine the nature of data transformation. In Sirius, these parameters wuold be the classification and mutual information parameters for the feature classification and mutual information data transformations, respectively. At this step, the system is also optionally able to export mutual information calculations for use in other contexts. \textbf{Transformed Data} In this step of the visualization pipeline, user interaction is classified by \cite{liu} as "Model Manipulation" and involves the augmentation of model parameters. Specifically, in Sirius, users can optionally adjust the significance threshold for backbone sparsification of the mutual information network from the rendered view. \textbf{Visual Mapping} Here, the dynamically thresheld mutual information matrix is mapped to a graph structure using a force-directed layout. Optionally, the graph structure can be exported for use in other applications. Users can augment the visual mapping parameter by selecting a different graph layout algorithm from the rendered views. \textbf{Visual Structure} Liu et. al. specify "Interactive Exploration" user interactions as augmentations to drawing parameters from the rendered views which determine the visual structure of the data to be displayed \cite{liu}. Specifically, in Sirius, this drawing parameter is the user interaction of hovering over an edge in the network graph to select a feature pair of interest. \textbf{View Transformation} When a user hovers over an edge corresponding to a bivariate data distribution, the view transformation step renders a new chart, of a type determined by the nature of the underlying data. Optionally, users can elect to export these charts outside of the dashboard structure to use in other contexts, e.g., as embedded interactive iframe elements in a standalone web article or report. \textbf{Rendered Views} The rendered views are the source for model manipulation and interactive exploration user interaction paradigms by way of augmenting model and drawing parameters. \textbf{User} The end user in the context of Sirius is data analysts performing cursory exploration of feature dependence for high-dimensional mixed data types. It is expected that these end users are familiar with all three types of user interaction paradigms: computation-centric, model manipulation, and interactive exploration.

\subsection{Views}

\begin{figure*}[htbp]
\centering
\includegraphics[width=17cm]{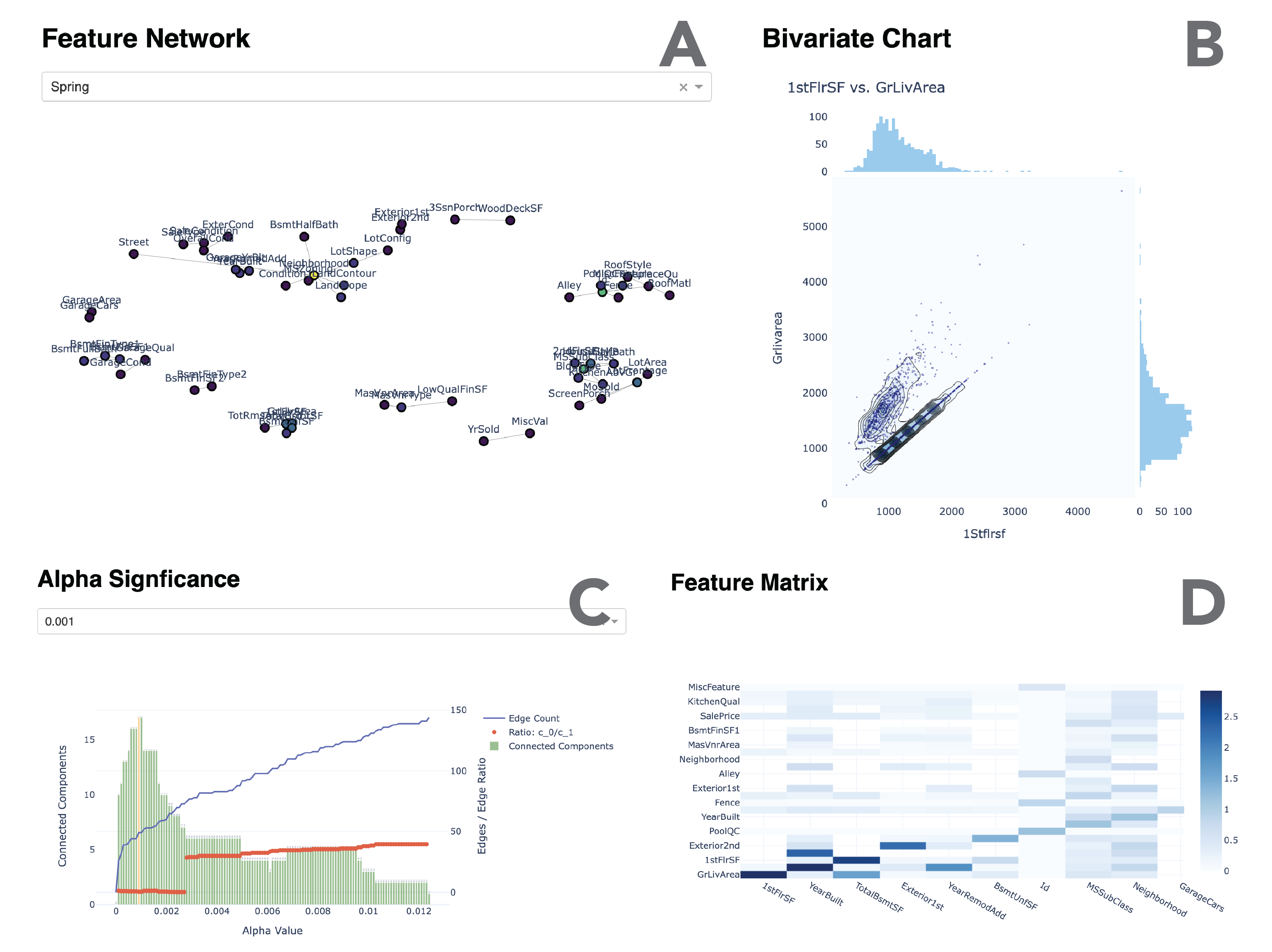}
\caption{\label{view_screenshots}
           A. The network graph of features is displayed, with a dropdown for selecting the network layout algorithm. B. When a user hovers over an edge in (A), the bivariate chart for that particular feature pair is shown in (B). C. The alpha significance chart shows metadata for signficance value, number of connected components, and the size ratio of node count for the first and second largest components. D. The matrix view shows the mutual information values of pairs of features, sparsified by the alpha significance value and sorted using hierarchical clustering.
}
\vspace{-14pt}
\end{figure*}

There are four main views in the Sirius interface: The network graph shown in Fig. \ref{view_screenshots} Panel A; the bivariate chart representation as shown in Fig. \ref{view_screenshots} Panel B; the alpha significance chart, seen in Fig. \ref{view_screenshots} Panel C; and the matrix view, shown in Fig. \ref{view_screenshots} Panel D. Clicking on an edge between two feature nodes in the network in (A) renders a bivariate chart in (B) of all records conditioned on the pair of features connected by the graph edge. The chart type rendered is determined by the data types of the two features being compared. These charts include beeswarm plots, heat maps, and 2D kernel density plots with scatter plot overlays. This information theoretic approach enables researchers to quickly view individual record data through statistically identified feature relationships of potential interest, with many tuning parameters for iterative customization in exploration.

The matrix representation of the thresheld feature network in (D) is, by nature, sparse, but is included for users as an alternative view in order to aid users in a logical transition from other matrix-based feature exploration tools such as SPLOMs \cite{scagnostics, scagexplorer}. Due to the sparse nature of the graph, we order features using hierarchical clustering in order to place similar groupings of features together, and additionally provide zoom support so that users can view neighborhoods of the matrix in greater detail.

When testing with real users, some data analysts expressed mistrust or confusion about the backbone sparsification paramater, prompting a request for the ability to adjust the alpha significance level via the rendered view. As a result, we have included a plot of the alpha significance value (C) as compared to 1) the total number of connected components in the graph; 2) the ratio of the largest component to the second-largest component (a common criteria when understanding network partitioning \cite{Serrano:2009:EMB}; and 3) the total number of edges in the resultant network graph. This chart is shown in Fig. \ref{view_screenshots} Panel C, and the view allows users to adjust manually the significance threshold using a dropdown. In this way, it is clearer for users to see how the graph becomes more densely connected as the significance value increases and the number of distinct components decreases.

\section{Case Study}\label{case_study}
For this case study, we focus on the \textbf{housing} data set. In particular, we review several of the user interactions available, and four example tasks served by the interface.

Consider a real estate marketing analyst, Maryam, who has received a new data set from a third party data broker. Maryam is hopeful that this data set will provide her with new insights into the homes in her area by \textbf{C2: Show dimension space and data item space}, but she is also skeptical: is this data clean? is it accurate? are there dangerous assumptions lurking under the surface? Before Maryam dives into complex data transformations and machine learning approaches, she just wants to \textit{look} at the data with her own eyes. She needs a tool that will \textbf{C3: Not distort raw data}. Unfortunately, this data is very wide (almost 100 different columns!), and viewing bivariate plots for every pair of features like in a scatterplot matrix would be a lot of plots to look at (well over 1,000 unique visualizations). Additionally, the data is a mix of discrete data like the name of the home's neighborhood, and continuous data like the home sale price. She needs to \textbf{C1: Support mixed data types} with her visual analysis. Luckily, Maryam has Sirius - thank goodness it's an \textbf{C5: Open source project}!. She uses the Python module to compute mutual information across all feature types, then launches the visualization dashboard. Maryam is happy that Sirius \textbf{C4: Ensures system modularity}, because now all of her mutual information computations, network graphs, and charts are available for use in her subsequent analysis workflow.

\begin{figure*}[htbp]
\centering
\includegraphics[width=17cm]{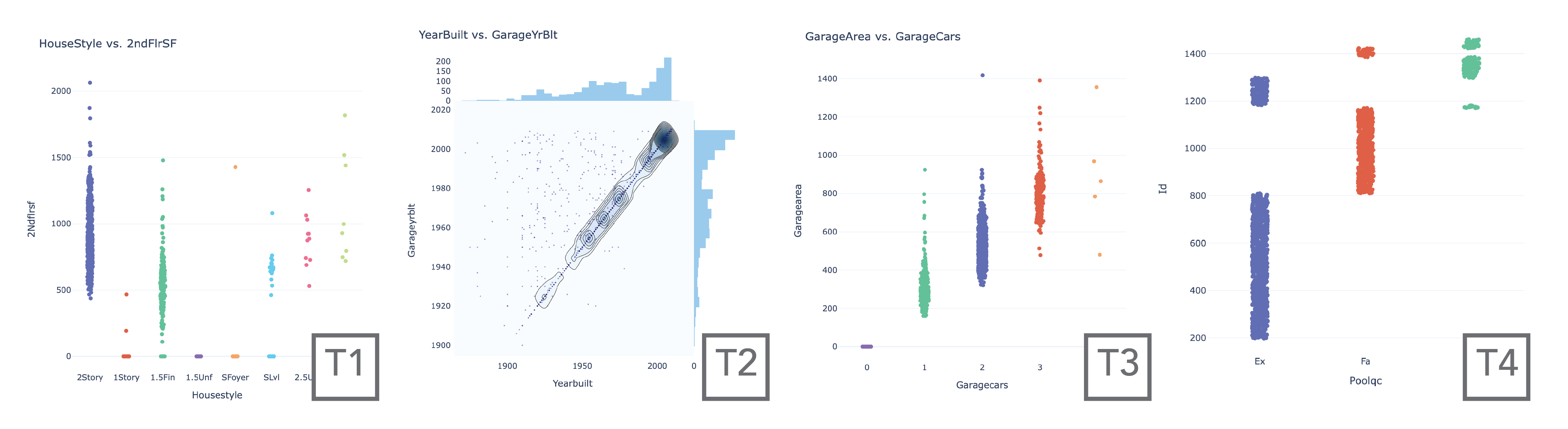}
\caption{\label{task_figure}
           \textbf{T1 Confirmatory Analysis} A bivariate plot showing House Style vs. Second Floor Square Footage, indicating a relationship between multi-level homes and higher second floor square footage. \textbf{T2 Suggested Feature Extraction} A bivariate plot is displayed showing ``Year Built" compared to the ``Garage Year Built". \textbf{T3 Identify Predictive Potential} Here we see a suggested feature plot from Sirius comparing ``Garage Cars" to ``Garage Area". \textbf{T4 Raise Anonymization Warnings} A screenshot from Sirius showing a bivariate plot of ``PoolQC" versus ``ID". ``PoolQC" is a quality measure which indicates 'Excellent' ('Ex'), 'Fair' ('Fa'), or Good ('Gd').
}
\end{figure*}

Revisiting the objectives detailed in Section \ref{objectives}, we consider 4 example tasks and demonstrate user interactions that could inform these tasks:
\begin{enumerate}
    \item \textbf{T1 Confirmatory Analysis}
    An important step in early analytics pipelines is confirmation of hypotheses and verification that data is recorded correctly. For this sort of confirmatory analysis, Sirius offers useful insight. For example, in the comparison shown in Fig. \ref{task_figure} Panel T1, we see a discrete-continuous plot of ``House Style" versus ``Second Floor Square Footage". As expected, the second floor square footage of homes is higher in 2-story and 2.5-story homes, followed by 1.5-story and split-level homes. Maryam is assured that these two variables are labeled correctly, and considers that if she needs a continuous variable for another prediction algorithm that is dependent upon the discrete variable style of home in her data set, she can consider the second floor square footage continuous variable as a potential candidate.
    
    \item \textbf{T2 Suggested Feature Extraction}
    Now Maryam is looking for potential feature pairings from which to extract new features of greater utility in her workflow. For this task, zooming in on particular network components may prove useful for finding extraction possibilities. As shown in Fig. \ref{task_figure} Panel T2, a strong linear relationship along the axis $x = y$ indicates that, for many homes, the garage was built at the same time as the house. Therefore, it may be of interest to an analyst to extract a new feature for the difference between ``Garage Year Built" and ``Year Built", in order to identify homes in which the garage was built much later than the house. Curiously, Maryam also sees some points where $x \leq y$, which could be errors to be targeted by data cleaning or follow-up, or may indicate homes where the house was demolished and rebuilt, but the garage remains from the old structure. Maryam resolves to call the data broker to ask, and perhaps look up some property records for the homes where $x \leq y$ before she moves forward with combining these two features into a new feature ``Garage Years Built After House".
    \item \textbf{T3 Identify Predictive Potential}
    Now Maryam considers another analytic task: the identification of potential predictive features. In Fig. \ref{task_figure} Panel T3, Maryam sees a suggested plot from Sirius showing ``Garage Cars" compared to ``Garage Area". Aha! This could be useful. It just so happens that from another data broker, Maryam's company has segmented satellite data of additional homes that are missing from this data set. It might be possible, using this known relationship, to predict the number of bay doors on the garages of those satellite data homes without having the street-level data or documentation to indicate the number of cars that the garage can accommodate. Maryam is happy, because she knows that customers on their website sometimes like to filter homes by how many cars the garage can fit, and the satellite data homes that are missing this information currently are dumped when that filter is applied. Now Maryam has an idea of a way to back-fill that data using a new feature relationship she has uncovered thanks to Sirius.
    \item \textbf{T4 Raise Anonymization Warnings}
    In a dataset that has been properly anonymized by randomly assigning IDs to records, Maryam expects that the ID number should not appear in the Sirius network graph, because considering the ID number as a continuous measure should not, statistically speaking, offer any benefit for mutual information, and should not be dependent on any other variables. However, in this case study data set, the ``ID" feature is present in the Sirius network graph! Maryam knows that this indicates that the ID value actually contains some information about the nature of the individual records. As shown in Fig. \ref{task_figure} Panel T4, there are complete ranges of the ``ID" number that correspond to the pool quality. This is a curious artifact which indicates that either: IDs are not randomly assigned, or possibly that the pool quality data is synthetically generated. Maryam resolves to investigate further by contacting homeowners about their pool quality and also asking the data broker how that data was collected or modeled.
\end{enumerate}

\section{Methods}

The Sirius data processing chronology is outlined as follows:
\begin{itemize}
  \item Classify features as discrete or continuous;
  \item Compute mutual information between each pair of features based on detected data type;
  \item Thin edges according to a `backbone' method which selects the most statistically unexpected edge weights for each node;
  \item Generate charts for each feature pair based on data type;
  \item Generate a network graph of features and weighted edges, and charts for each pair of features to be displayed on edge selection.\end{itemize}
  
  In this system, a network graph is generated in which nodes represent features, and edges are weighted between features by mutual information score. Edge display is thresholded dynamically, informed by a `backbone' graph thinning method commonly employed in network science to extract the critical structures of a complex graph\cite{Bagrow:2015:RMS,Serrano:2009:EMB}.

\subsection{Data}
Prior to visualization, data must be processed in order to generate mutual information scores for all feature pairs. For this step, users would specify a tabular data file location and, optionally, set parameters for feature classification as discrete or continuous, and set algorithm parameters for mutual information calculation.

We provide three domain-specific example data sets with Sirius, and discuss one such data set as a case study in Section \ref{case_study}.

Example publicly available data sets used with this tool include (in order of record volume):
\begin{itemize}
\item A \textbf{healthcare} data set related to intensive care unit (ICU) mortality, comprised of 186 features: a mixture of demographic information, lab results, and risk assessments (e.g ``Blood Glucose", ``Weight", and ``Apache 2 Diagnosis") from 91,713 records of admitted ICU patients \cite{Lee:2020:WID}.
\item A \textbf{groceries} data set comprised of 169 features: unique items in a grocery store (e.g., ``Vegetables", ``Cleaning Supplies", and ``Specialty Meats"), with boolean values (corresponding to `purchased' or `not purchased') from 9,835 records of market basket transactions \cite{Nasrullah:2019:GMB}. This data set is unique in our examples, in that it only contains a binary response: all features only have 1 of 2 possible answers, and this data set is included primarily to demonstrate that the mutual information feature network technique can also be employed for data sets of homogenous data types, e.g., all discrete or all continuous features, and the resultant networks can be compared to other feature networks such as the association network.
\item The Ames \textbf{housing} data set is comprised of 81 features of mixed data types for 1,460 records of homes in Ames, Iowa \cite{DeCock:2019:HPA}. It includes physical characteristics of homes such as features and materials, as well as qualitative assessments and sale information. This is the dataset that we use for the case study with task examples here in Section \ref{case_study}.
\end{itemize}
The resultant network graphs for these data sets group features with statistically significant dependence, with potential applications for: clinicians using data similar to the ICU mortality data set; marketers with data resembling the groceries data set; or real estate agents or tax assessors examining data paralleling the housing data set. This indicates promise for other real-world applications which require comparisons across a large number of features with differing data types.

A synthetic data set is also included in the package, with 34 features of mixed data types for 2,400 artificial records. feature names correspond to statistical properties of synthesized data. This data is not discussed here, as it is primarily for interface testing and demonstration purposes.

\subsection{Calculating mutual information}\label{calc_mi}

An \textbf{edge list} is obtained by taking the upper or lower triangle of an undirected (symmetric) $n$ x $n$ feature matrix, excluding the diagonal matrix identity. The resulting edge list has a length $\frac{n^2-n}{2}$, such that all feature nodes are connected once, prior to sparsification.

Mutual information ($I$) is computed between each pair of features ($U,V$), and listed as the weight of the corresponding edge. The default behavior of the application is to calculate mutual information only across records for which there is a non-null value for both features. Thus, if there are no records for which $U$ and $V$ are non-null, $I_{U;V} = 0$. Users might opt to consider $r = $ \texttt{null} as an additional discrete `category' (for example, if the lack of a response might be considered informative for the domain application), or fill missing continuous values with the range minimum or median.

\begin{figure}
\centering
\includegraphics[width=0.5\linewidth]{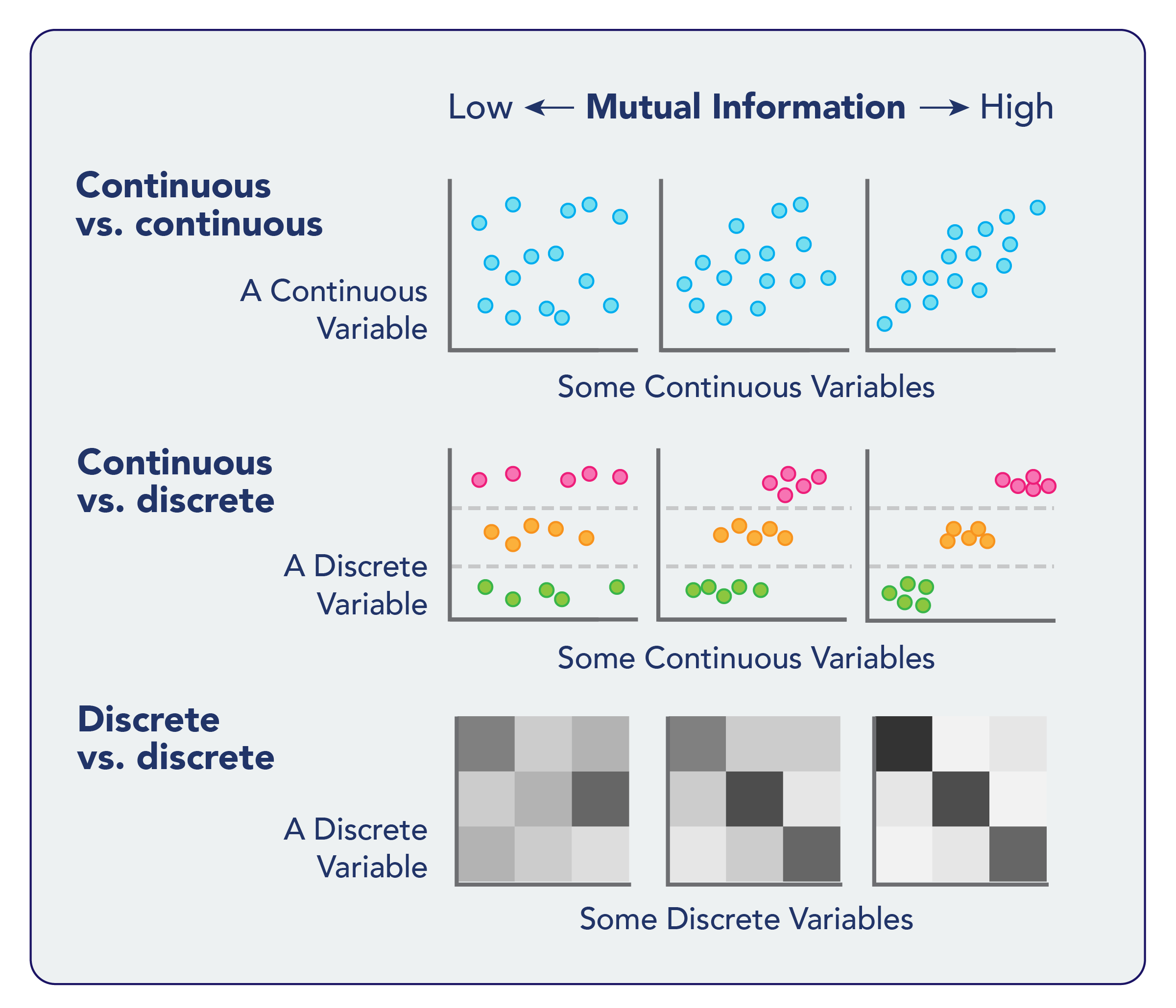}
\caption{\label{branching}
           The mutual information algorithm is chosen based on the feature types.
}
\end{figure}

\textbf{Discrete/Discrete Data Type Pairs}
For a pair of discrete data type features, mutual information is calculated using probability mass functions across all non-null ${i\in U}$ and ${j\in V}$. The mutual information formula for discrete-discrete pairs is given as: 

\begin{equation}
I_{U;V} = \sum_{i\in U}\sum_{j\in V}=P(i,j)+log\frac{P(i,j)}{P(i)*P(j)}
\end{equation}

\textbf{Continuous/Continuous Data Type Pairs}
Converting this method from probability mass functions to probability density functions for a pair of continuous data type features yields the formula: 

\begin{equation}
I_{U;V} = \int_{U} \int_{V} P_{(U,V)}(i,j)+log\frac{P_{(U,V)}(i,j)}{P_U(i)*P_V(j)}\ di\ dj
\end{equation}

where $P_{(U,V)}$ is the joint probability density function of $U$ and $V$, and $P_U$,$P_V$ are the marginal probability density functions of $U$ and $V$, respectively\cite{Kraskov:2004:EMI}. 

\textbf{Discrete/Continuous Data Type Pairs}
A heterogenous pairing of data types employs a nearest neighbor mutual information regression method\cite{Ross:2014:MIB,Pedregosa:2011:SLM}. Discrete features with $r$ responses are converted from a 1D array of length $m$ to a sparse matrix of shape ($r$, $m$), with columns for each unique response and values of 0 or 1 for all records $k$. Nearest neighbor approximations of mutual information in a sample population are less susceptible to variance due to parameterization than traditional binning methods; e.g., adjusting the neighbor $N$ count for each point $x$ does not influence the variance in mutual information as significantly as adjusting the bandwidth of a Gaussian kernel\cite{Ross:2014:MIB}. 

\begin{equation}
I_{(U,V)}= 	\langle I_i\rangle = \psi(N_x) - \langle\psi(N_x)\rangle + \psi(k) - \psi(m)
\end{equation}

Following the behavior prescribed by the libraries used as detailed in Section \ref{technique}, ${[{u\in U},{v\in V}]}: {I_{(u;v)} \geq 0}$, as mutual information cannot be negative, since it is a measure of dependence and zero corresponds to complete independence\cite{Pedregosa:2011:SLM,Kraskov:2004:EMI}.

\subsection{Network Graphing}\label{network_methods}

\begin{figure}
\centering
\includegraphics[width=\linewidth]{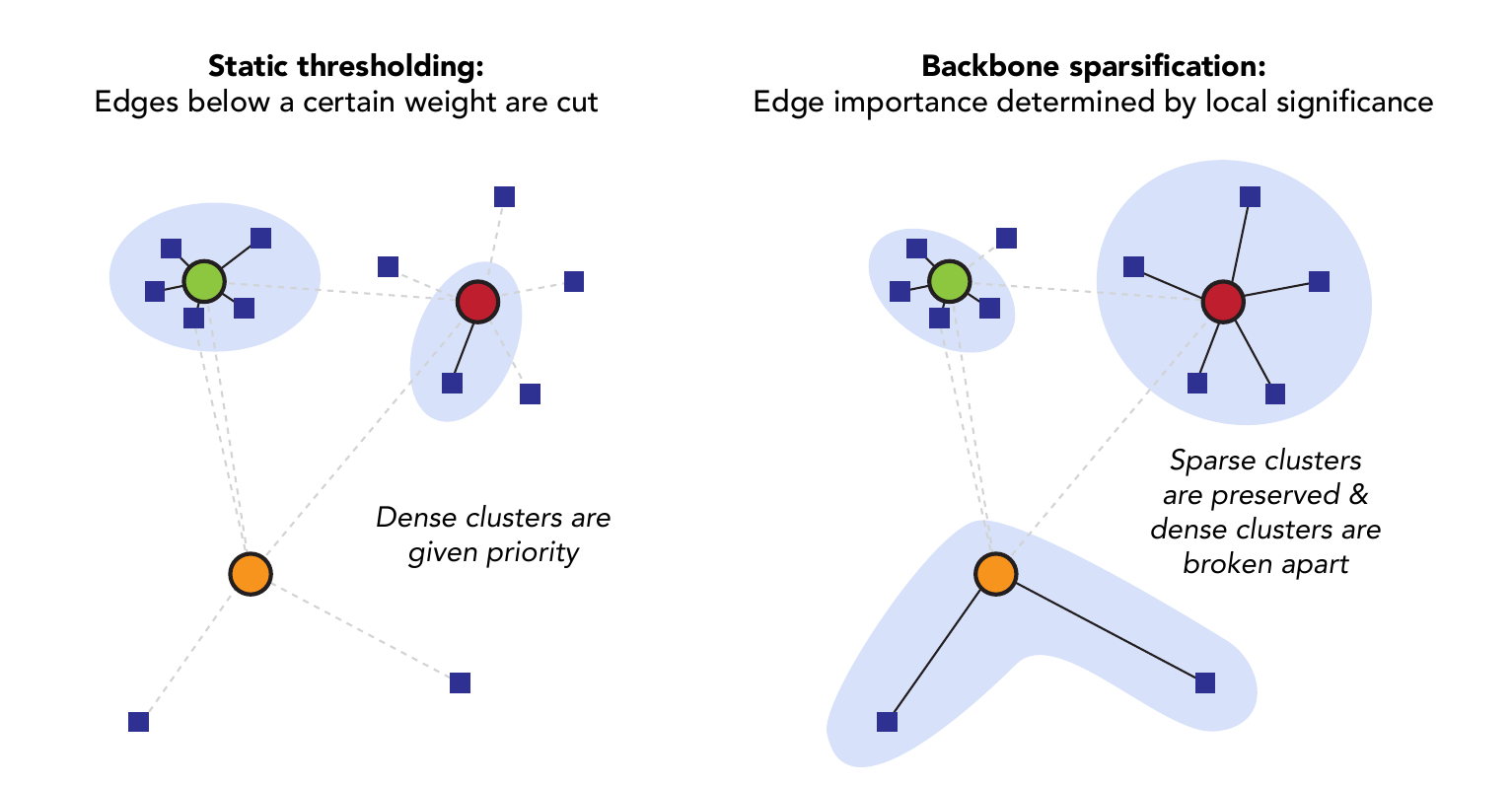}
\caption{\label{thresholding}
           The value of the backbone method as a dynamic thresholding sparsification approach, as compared to static thresholding, is that backbone sparsification is network-aware. In real-world data contexts, often certain features will be densely connected to their neighbors with respect to mutual information, because they are near-identical matches for one another. For example, in a healthcare setting, ``hemoglobin\_apache" might simply be a value that is computed from ``hemoglobin\_1a" and ``hemoglobin\_1b". Therefore, thresholding the network graph by the mutual information value alone at a static threshold would result in the visual prioritization of extremely dense clusters, but miss the statistically interesting feature relationships. We use backbone thresholding to assign an alpha significance value to each edge based on the relative weights of other edges in the network, in order to sparsify the matrix in a more graph-aware manner.
}
\end{figure}

Sparsification of the resulting feature matrix allows for a narrowing of the exploratory visualization space in a structurally aware manner, as shown in Figure \ref{thresholding}. In this technique, edge thinning is performed through a `backbone' method, which preserves links with high statistical significance and removes all others \cite{Bagrow:2015:RMS,Serrano:2009:EMB}. The equation is outlined as follows:

\begin{equation}
\alpha_{ij} = 1 - (k - 1) \int_{0}^{P_{ij}} (1-x)^{k-2} dx < \alpha
\end{equation}

where $\alpha_{ij}$ is the statistical significance of a given edge, $k$ is the node degree (the number of nodes that this node is connected to), $P_{ij}$ is the probability of an edge with this weight $x$, and $\alpha$ is the significance threshold. 

\begin{figure}
\centering
\includegraphics[width=\linewidth]{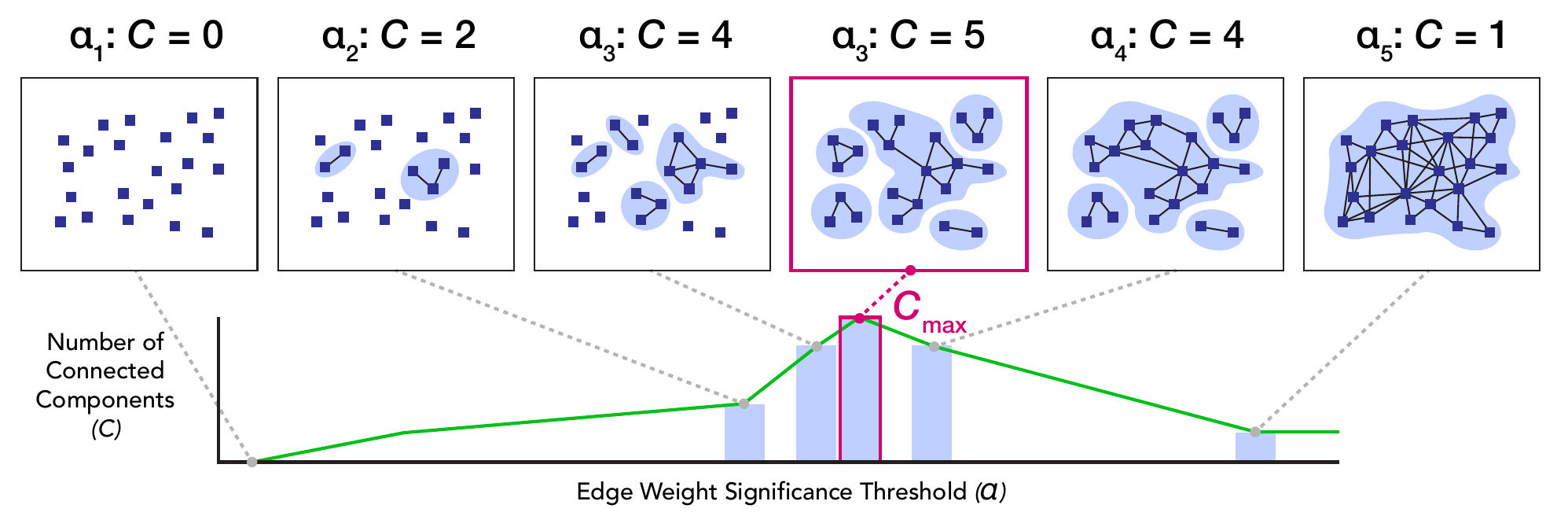}
\caption{\label{alpha_explanation}
           When the alpha significance is at 0, no nodes are connected, and thus the number of components $C$ is 0 ($\alpha_{1}$). As the alpha threshold for the network is adjusted upwards, the number of edges increases, creating more connected components. At a certain alpha threshold (here, $\alpha_{3}$), the number of connected components in the graph reaches a maximum (here, $n_C$ = 5 at $\alpha_{3}$), before decreasing as discrete components become connected to one another, finally resulting in a single connected component encompassing the entire graph ($\alpha_{5}$).
}
\end{figure}

In this way, there is not a specific mutual information threshold set for the entire network, but rather each node is evaluated individually, and its edges are preserved or thinned according to how relatively important an edge is to that node specifically. An alpha threshold is chosen for which the number of graph components $C$ is maximal, which provides for unique 'constellations' in the visualization, as shown in Fig \ref{alpha_explanation}. 

Network statistics such as adjacency and component count are calculated using NetworkX. Graph positioning is also calculated using NetworkX\cite{NetworkX:2014:NWX}. Layout parameters can be adjusted within the graph layout function. Default behavior is to employ the Fruchterman Reingold spring layout, a commonly used force-directed positioning algorithm in which, in this mutual information network use case, highly dependent features are more closely positioned than features which are more independent from one another\cite{Fruchterman:1991:GDF}. However, we also provide users with the ability to toggle to the Kamada-Kawai layout, as another view for exporation of the same edge list, just with different positioning of nodes in X,Y coordinate space.

Notably, in order to enable user interaction on edge hover, we have also computed a ``shadow graph" of the rendered graph wherein a hidden node lies at the midpoint of each edge and corresponds to some bivariate feature coupling. Therefore, functionally what is happening when a user interacts with an edge in the network graph in the left-hand panel, is that a hidden node is activated which then makes a function call to render the associated bivariate chart in the right-hand panel.

\section{Discussion}
There are a number of ways in which we see this novel approach as extensible through the addition of modules, and more complex analysis of the mutual information feature network itself:

\subsection{Charts}
We have selected three chart types (heatmap, beeswarm plot, and 2D kernel density) for each of the three permutations of data type pairing supported by Sirius. However, other visual encodings could be investigated. For example, grouped bar charts may be a more intuitive alternative to heatmaps for some users when comparing two discrete features; boxplots, violin plots, or bean plots are valid alternative visual encodings for pairings of discrete and continuous data types. These alternatives could be provided as parameter settings or toggles in future iterations of user interface design.

\subsection{Advanced Network Analysis Measures}
We are careful here not to compare feature networks to real-world networks (i.e., social networks, ecological systems, power grids). The network visualization is used simply as a relational graph. It remains to be seen whether the mutual information feature network (or association networks or gene correlation networks) are subject to the same properties as real-world networks, e.g., the `friendship paradox' (in which the average node is connected to nodes with a higher average number of connections), or eligible for traditional network study such as percolation theory or community detection \cite{Feld:1991:WYF}. Meta-analysis of the properties of various mutual information feature networks could yield interesting insight into organic feature relationship behavior, as compared to null models of synthetically generated data. For example, the connectivity or graph spectra of real mutual information networks could be fundamentally different from feature networks of synthesized high-dimensional data, which may help data scientists to more easily identify cases of data fraud or imputation through examination of anomalous feature relationships \cite{spider_fraud}.

\subsection{Confounding Variables}
The mutual information feature network, as with other feature networks used for feature selection and data mining, is concerned only with edges between pairs of features, not multidimensional interactions between more than two features. feature networks fall short of addressing statistical concerns such as Simpson's Paradox, in which unidentified tertiary conditions render summary conclusions invalid \cite{Berman:2012:SPC}. Future research could involve topological analysis of feature networks to identify confounding variables and generate multivariate visualizations upon interaction with network graph faces (contained by 3 or more vertices) through the use of simplicial complexes \cite{jonsson2008simplicial}.

\subsection{Temporal Evolution}
Functionality could also be implemented for temporal animation of mutual information feature networks, illustrating how feature dependence relationships strengthen or weaken over time. For example, cholesterol values may have been more informative for predictive wellness modeling prior to the widespread usage of cholesterol-lowering drugs; square footage of homes may have been more closely associated with sale price before the introduction of luxury `tiny homes' in a neighborhood. Temporal animation of mutual information feature networks might bring the dynamics of changing feature relationships to the forefront in an exploratory analysis.

\section{Conclusions}
Initial observations of the mutual information feature network show promise for the method as a new technique for exploratory analysis. Parameterization preserves customizability for specific data structures and domain applications. Further statistical measures of resultant network structures offer potential avenues for insight into more complex feature relationships in high-dimensional data analysis.

\section{Acknowledgments}
This work was supported in part by funding from Massachusetts Mutual Life Insurance Company, through the establishment of the MassMutual Center of Excellence in Complex Systems and Data Science at the University of Vermont. Thanks also to David Rushing Dewhurst for aiding our understanding of mutual information and kernel density estimation; Mark Meyers, Adam Fox, and Sears Merritt at MassMutual Data Science for domain expertise; and Juniper Lovato, Melissa Rubinchuk, and Alex Woodward at the University of Vermont for coordinating and supporting this institutional research partnership. We have additional gratitude for Michelle Borkin at Northeastern, who advised on subsequent versions of this manuscript.

The authors are grateful to the anonymous reviewers of EuroVis 2020 short papers; TVCG / IEEE VIS 2020 and 2021; and ACM IUI 2020; for their thorough and thoughtful feedback, which has been incorporated into this pre-print where possible.
\bibliographystyle{unsrtnat}
\bibliography{bibliography}

\end{document}